\documentstyle[12pt]{article}

\advance \topmargin by -\headheight
\advance \topmargin by -\headsep
     
\evensidemargin \oddsidemargin
\begin{document}

\topmargin -.6in
\def\nonu{\nonumber}
\def\rf#1{(\ref{eq:#1})}
\def\lab#1{\label{eq:#1}} 
\def\br{\begin{eqnarray}}
\def\er{\end{eqnarray}}
\def\be{\begin{equation}}
\def\ee{\end{equation}}
\def\0{\nonumber}
\def\lb{\lbrack}
\def\rb{\rbrack}
\def\({\left(}
\def\){\right)}
\def\v{\vert}
\def\bv{\bigm\vert}
\def\lskip{\vskip\baselineskip\vskip-\parskip\noindent}
\relax
\newcommand{\nit}{\noindent}
\newcommand{\ct}[1]{\cite{#1}}
\newcommand{\bi}[1]{\bibitem{#1}}
\def\a{\alpha}
\def\b{\beta}
\def\ca{{\cal A}}
\def\cm{{\cal M}}
\def\cn{{\cal N}}
\def\cf{{\cal F}}
\def\d{\delta} 
\def\D{\Delta}
\def\eps{\epsilon}
\def\g{\gamma}
\def\G{\Gamma}
\def\grad{\nabla}
\def\h{ {1\over 2}  }
\def\hc{\hat{c}}
\def\hd{\hat{d}}
\def\hg{\hat{g}}
\def\hp{ {+{1\over 2}}  }
\def\hm{ {-{1\over 2}}  }
\def\k{\kappa}
\def\l{\lambda}
\def\L{\Lambda}
\def\lg{\langle}
\def\m{\mu}
\def\n{\nu}
\def\o{\over}
\def\om{\omega}
\def\O{\Omega}
\def\p{\phi}
\def\pa{\partial}
\def\pr{\prime}
\def\ra{\rightarrow}
\def\rh{\rho}
\def\rg{\rangle}
\def\s{\sigma}
\def\t{\tau}
\def\th{\theta}
\def\ti{\tilde}
\def\wti{\widetilde}
\def\inte{\int dx }
\def\xb{\bar{x}}
\def\yb{\bar{y}}

\def\tr{\mathop{\rm tr}}
\def\Tr{\mathop{\rm Tr}}
\def\partder#1#2{{\partial #1\over\partial #2}}
\def\ds{{\cal D}_s}
\def\wtwo{{\wti W}_2}
\def\lie{{\cal G}}
\def\alie{{\widehat \lie}}
\def\dlie{{\cal G}^{\ast}}
\def\elie{{\widetilde \lie}}
\def\edlie{{\elie}^{\ast}}
\def\hlie{{\cal H}}
\def\wlie{{\widetilde \lie}}

\def\rlx{\relax\leavevmode}
\def\inbar{\vrule height1.5ex width.4pt depth0pt}
\def\IZ{\rlx\hbox{\sf Z\kern-.4em Z}}
\def\IR{\rlx\hbox{\rm I\kern-.18em R}}
\def\IC{\rlx\hbox{\,$\inbar\kern-.3em{\rm C}$}}
\def\one{\hbox{{1}\kern-.25em\hbox{l}}}

\def\PRL#1#2#3{{\sl Phys. Rev. Lett.} {\bf#1} (#2) #3}
\def\NPB#1#2#3{{\sl Nucl. Phys.} {\bf B#1} (#2) #3}
\def\NPBFS#1#2#3#4{{\sl Nucl. Phys.} {\bf B#2} [FS#1] (#3) #4}
\def\CMP#1#2#3{{\sl Commun. Math. Phys.} {\bf #1} (#2) #3}
\def\PRD#1#2#3{{\sl Phys. Rev.} {\bf D#1} (#2) #3}
\def\PLA#1#2#3{{\sl Phys. Lett.} {\bf #1A} (#2) #3}
\def\PLB#1#2#3{{\sl Phys. Lett.} {\bf #1B} (#2) #3}
\def\JMP#1#2#3{{\sl J. Math. Phys.} {\bf #1} (#2) #3}
\def\PTP#1#2#3{{\sl Prog. Theor. Phys.} {\bf #1} (#2) #3}
\def\SPTP#1#2#3{{\sl Suppl. Prog. Theor. Phys.} {\bf #1} (#2) #3}
\def\AoP#1#2#3{{\sl Ann. of Phys.} {\bf #1} (#2) #3}
\def\PNAS#1#2#3{{\sl Proc. Natl. Acad. Sci. USA} {\bf #1} (#2) #3}
\def\RMP#1#2#3{{\sl Rev. Mod. Phys.} {\bf #1} (#2) #3}
\def\PR#1#2#3{{\sl Phys. Reports} {\bf #1} (#2) #3}
\def\AoM#1#2#3{{\sl Ann. of Math.} {\bf #1} (#2) #3}
\def\UMN#1#2#3{{\sl Usp. Mat. Nauk} {\bf #1} (#2) #3}
\def\FAP#1#2#3{{\sl Funkt. Anal. Prilozheniya} {\bf #1} (#2) #3}
\def\FAaIA#1#2#3{{\sl Functional Analysis and Its Application} {\bf #1} (#2)
#3}
\def\BAMS#1#2#3{{\sl Bull. Am. Math. Soc.} {\bf #1} (#2) #3}
\def\TAMS#1#2#3{{\sl Trans. Am. Math. Soc.} {\bf #1} (#2) #3}
\def\InvM#1#2#3{{\sl Invent. Math.} {\bf #1} (#2) #3}
\def\LMP#1#2#3{{\sl Letters in Math. Phys.} {\bf #1} (#2) #3}
\def\IJMPA#1#2#3{{\sl Int. J. Mod. Phys.} {\bf A#1} (#2) #3}
\def\AdM#1#2#3{{\sl Advances in Math.} {\bf #1} (#2) #3}
\def\RMaP#1#2#3{{\sl Reports on Math. Phys.} {\bf #1} (#2) #3}
\def\IJM#1#2#3{{\sl Ill. J. Math.} {\bf #1} (#2) #3}
\def\APP#1#2#3{{\sl Acta Phys. Polon.} {\bf #1} (#2) #3}
\def\TMP#1#2#3{{\sl Theor. Mat. Phys.} {\bf #1} (#2) #3}
\def\JPA#1#2#3{{\sl J. Physics} {\bf A#1} (#2) #3}
\def\JSM#1#2#3{{\sl J. Soviet Math.} {\bf #1} (#2) #3}
\def\MPLA#1#2#3{{\sl Mod. Phys. Lett.} {\bf A#1} (#2) #3}
\def\JETP#1#2#3{{\sl Sov. Phys. JETP} {\bf #1} (#2) #3}
\def\JETPL#1#2#3{{\sl  Sov. Phys. JETP Lett.} {\bf #1} (#2) #3}
\def\PHSA#1#2#3{{\sl Physica} {\bf A#1} (#2) #3}
\def\PHSD#1#2#3{{\sl Physica} {\bf D#1} (#2) #3}
\begin{center}
{\large\bf    Permutability of Backlund Transformation for $N=1$ Supersymmetric Sinh-Gordon  \footnote{ key words:Backlund Transformation,
Supersymmetric sinh-Gordon,
non linear superposition formula \\Pacs 02.30.Ik}}
\end{center}
\normalsize
\vskip .4in

\begin{center}
J.F. Gomes\footnote{corresponding author jfg@ift.unesp.br},  L.H. Ymai and A.H. Zimerman 

\par \vskip .1in \noindent
Instituto de F\'{\i}sica Te\'{o}rica-UNESP\\
Rua Pamplona 145\\
fax (55)11 31779080\\
01405-900 S\~{a}o Paulo, Brazil
\par \vskip .3in

\end{center}
\baselineskip=1cm
\begin{abstract}
The permutability of two Backlund transformations  is employed  
to construct a non linear superposition formula  to generate  a class of solutions for the $N=1$ super sinh-Gordon model.
\end{abstract}

Backlund transformations (BT) relating two different soliton solutions  are known to be characteristic 
of certain  class   of nonlinear  equations.  
A  remarkable consequence is that from a particular soliton solution, a second solution  can be generated by  Backlund transformation.  
This second  solution, in turn,  generates a third one  and  
such structure allows  to construct conditions for  the permutability 
of two sequences of BT.
 
The study of   superposition principle for soliton solutions  of the sine-Gordon equation 
was employed to show  that the order of two BT is, in fact, irrelevant \cite{fordy}. 
Such condition became known as the {\it permutability theorem} as was applied  
 to  the    KdV  and for the $N=1$ super KdV  equations  
in  \cite{wahlquist} and in \cite{liu} respectively.

Soliton solutions for the $N=1$ super sine-Gordon were obtained  in \cite{gram} 
from  the super Hirota's  formalism and in \cite{ymaipla} using  dressing transformation 
and vertex operators.     Backlund solutions  for super mKdV were considered in \cite{liu2}. 
In this paper we use the superfield approach for the BT as proposed in \cite{chai} 
to derive a closed  algebraic   superposition formula for soliton solutions of 
the $N=1$ super sinh-Gordon  model
assuming that  two sucessive BT 
commute.

The model is described by the following equation of motion written within the superfield formalism \cite{chai}
\begin{eqnarray}
D_xD_t\Phi=2i\sinh\Phi,\label{1}
\end{eqnarray}
where the bosonic superfield  $\Phi$ is given in components by
\begin{eqnarray}
\Phi=\phi+\theta_1\bar{\psi}+i\theta_2\psi-\theta_1\theta_2 2i\sinh\phi,\label{2}
\end{eqnarray}
where $\theta_1$ and $\theta_2$ are Grassmann variables (i.e. $\theta_1^2 = \theta_2^2 = 0$ and $\theta_1 \theta_2 +\theta_2 \theta_1=0$ )
The superderivatives
\begin{eqnarray}
D_x=\partial_{\theta_1}+\theta_1\partial_x, \qquad
D_t=\partial_{\theta_2}+\theta_2\partial_t,\label{3}
\end{eqnarray}
satisfy
\begin{eqnarray}
D_x^2=\partial_x, \qquad D_t^2=\partial_t, \qquad D_xD_t=-D_tD_x.\label{4}
\end{eqnarray}
The  Backlund transformation for eqn. (\ref{1}) is given by \cite{chai}
\begin{eqnarray}
D_x(\Phi_0-\Phi_1)&=&-\frac{4i}{\beta_1}f_{0,1}\cosh\left(\frac{\Phi_0+\Phi_1}{2}\right),  \label{5a} \\
D_t(\Phi_0+\Phi_1)&=&2\beta_1
f_{0,1}\cosh\left(\frac{\Phi_0-\Phi_1}{2}\right),\label{5b}
\end{eqnarray}
where $\b_1$ is an arbitrary parameter (spectral parameter) and the auxiliary fermionic superfield $f_{0,1}$ (i.e. $f_{0,1}^2 = 0$)
\begin{eqnarray}
f_{0,1}=f_1^{(0,1)}+\theta_1b_1^{(0,1)}+\theta_2b_2^{(0,1)}+\theta_1\theta_2f_2^{(0,1)},\label{6}
\end{eqnarray}
satisfy
\begin{eqnarray}
D_xf_{0,1}=\frac{2i}{\beta_1}\sinh\left(\frac{\Phi_0+\Phi_1}{2}\right), \qquad 
D_tf_{0,1}=\beta_1\sinh\left(\frac{\Phi_0-\Phi_1}{2}\right). \label{7a} 
\end{eqnarray}
Consider now two  successive  Backlund transformations.  The first one involving superfields $\Phi_0$ and $\Phi_1$   and the parameter $\b_1$ whilst the second involves $\Phi_1$ and $\Phi_3$ with  $\b_2$.  The  {\it Permutability theorem} states that the order in which such Backlund transformations are employed is irrelevant, i.e.  we might as well consider  the first involving $\Phi_0$ and $\Phi_2$ with $\b_1$ followed by a second, involving $\Phi_2$ and $\Phi_3$ with $\b_2$.  Similar to (\ref{5a})  we have,
\begin{eqnarray}
D_x(\Phi_0-\Phi_1)&=&-\frac{4i}{\beta_1}f_{0,1}\cosh\left(\frac{\Phi_0+\Phi_1}{2}\right),\label{Bx01}\\
D_x(\Phi_1-\Phi_3)&=&-\frac{4i}{\beta_2}f_{1,3}\cosh\left(\frac{\Phi_1+\Phi_3}{2}\right),\label{Bx13}\\
D_x(\Phi_0-\Phi_2)&=&-\frac{4i}{\beta_2}f_{0,2}\cosh\left(\frac{\Phi_0+\Phi_2}{2}\right),\label{Bx02}\\
D_x(\Phi_2-\Phi_3)&=&-\frac{4i}{\beta_1}f_{2,3}\cosh\left(\frac{\Phi_2+\Phi_3}{2}\right).\label{Bx23}
\end{eqnarray}
and from (\ref{7a}),
\begin{eqnarray}
D_xf_{0,1}&=&\frac{2i}{\beta_1}\sinh\left(\frac{\Phi_0+\Phi_1}{2}\right),\label{fx01}\\
D_xf_{1,3}&=&\frac{2i}{\beta_2}\sinh\left(\frac{\Phi_1+\Phi_3}{2}\right),\label{fx13}\\
D_xf_{0,2}&=&\frac{2i}{\beta_2}\sinh\left(\frac{\Phi_0+\Phi_2}{2}\right),\label{fx02}\\
D_xf_{2,3}&=&\frac{2i}{\beta_1}\sinh\left(\frac{\Phi_2+\Phi_3}{2}\right).\label{fx23}
\end{eqnarray}

The equality of the sum of  equations (\ref{Bx01}) and  (\ref{Bx13})  with the sum of 
 (\ref{Bx02}) and (\ref{Bx23}) yields the following relation
 \begin{eqnarray}
&&\frac{1}{\beta_1}f_{0,1}\cosh\left(\frac{\Phi_0+\Phi_1}{2}\right)+\frac{1}{\beta_2}f_{1,3}\cosh\left(\frac{\Phi_1+\Phi_3}{2}\right)\nonumber\\
&&=\frac{1}{\beta_2}f_{0,2}\cosh\left(\frac{\Phi_0+\Phi_2}{2}\right)+\frac{1}{\beta_1}f_{2,3}\cosh\left(\frac{\Phi_2+\Phi_3}{2}\right).\label{Rx}
\end{eqnarray}
Analogously from (\ref{5b}) we find
\begin{eqnarray}
D_t(\Phi_0+\Phi_1)&=&2\beta_1
f_{0,1}\cosh\left(\frac{\Phi_0-\Phi_1}{2}\right),\label{Bt01}\\
D_t(\Phi_1+\Phi_3)&=&2\beta_2
f_{1,3}\cosh\left(\frac{\Phi_1-\Phi_3}{2}\right),\label{Bt13}\\
D_t(\Phi_0+\Phi_2)&=&2\beta_2
f_{0,2}\cosh\left(\frac{\Phi_0-\Phi_2}{2}\right),\label{Bt02}\\
D_t(\Phi_2+\Phi_3)&=&2\beta_1
f_{2,3}\cosh\left(\frac{\Phi_2-\Phi_3}{2}\right).\label{Bt23}
\end{eqnarray}
Equating  now the difference of the first two, (\ref{Bt01}) and  (\ref{Bt13})  and the last two equations, (\ref{Bt02}) and  (\ref{Bt23}) we get
\begin{eqnarray}
&&\beta_1f_{0,1}\cosh\left(\frac{\Phi_0-\Phi_1}{2}\right)-\beta_2f_{1,3}\cosh\left(\frac{\Phi_1-\Phi_3}{2}\right)\nonumber\\
&&=\beta_2f_{0,2}\cosh\left(\frac{\Phi_0-\Phi_2}{2}\right)-\beta_1f_{2,3}\cosh\left(\frac{\Phi_2-\Phi_3}{2}\right).\label{Rt}
\end{eqnarray}
Solving (\ref{Rx}) and (\ref{Rt}), for $f_{1,3}$ and $f_{2,3}$, we get
\begin{eqnarray}
f_{1,3}&=&\Lambda_{1,3}^{(1)}f_{0,1}+\Lambda_{1,3}^{(2)}f_{0,2},\label{f13}\\
f_{2,3}&=&\Lambda_{2,3}^{(1)}f_{0,1}+\Lambda_{2,3}^{(2)}f_{0,2},\label{f23}
\end{eqnarray}
where the coefficients $\L$ are given as
\begin{eqnarray}
\Lambda_{1,3}^{(1)}=-\beta_1\beta_2\frac{\left[\cosh\left(\frac{\Phi_0+\Phi_1}{2}\right)\cosh\left(\frac{\Phi_2-\Phi_3}{2}\right)+\cosh\left(\frac{\Phi_0-\Phi_1}{2}\right)\cosh\left(\frac{\Phi_2+\Phi_3}{2}\right)\right]}{\left[\cosh\left(\frac{\Phi_2-\Phi_3}{2}\right)\cosh\left(\frac{\Phi_1+\Phi_3}{2}\right)\beta_1^2-\cosh\left(\frac{\Phi_1-\Phi_3}{2}\right)\cosh\left(\frac{\Phi_2+\Phi_3}{2}\right)\beta_2^2\right]},\nonumber
\end{eqnarray}
\begin{eqnarray}
\Lambda_{1,3}^{(2)}=\frac{\left[\cosh\left(\frac{\Phi_0+\Phi_2}{2}\right)\cosh\left(\frac{\Phi_2-\Phi_3}{2}\right)\beta_1^2+\cosh\left(\frac{\Phi_0-\Phi_2}{2}\right)\cosh\left(\frac{\Phi_2+\Phi_3}{2}\right)\beta_2^2\right]}{\left[\cosh\left(\frac{\Phi_2-\Phi_3}{2}\right)\cosh\left(\frac{\Phi_1+\Phi_3}{2}\right)\beta_1^2-\cosh\left(\frac{\Phi_1-\Phi_3}{2}\right)\cosh\left(\frac{\Phi_2+\Phi_3}{2}\right)\beta_2^2\right]},\nonumber
\end{eqnarray}
\begin{eqnarray}
\Lambda_{2,3}^{(1)}=-\frac{\left[\cosh\left(\frac{\Phi_0-\Phi_1}{2}\right)\cosh\left(\frac{\Phi_1+\Phi_3}{2}\right)\beta_1^2+\cosh\left(\frac{\Phi_0+\Phi_1}{2}\right)\cosh\left(\frac{\Phi_1-\Phi_3}{2}\right)\beta_2^2\right]}{\left[\cosh\left(\frac{\Phi_2-\Phi_3}{2}\right)\cosh\left(\frac{\Phi_1+\Phi_3}{2}\right)\beta_1^2-\cosh\left(\frac{\Phi_1-\Phi_3}{2}\right)\cosh\left(\frac{\Phi_2+\Phi_3}{2}\right)\beta_2^2\right]},\nonumber
\end{eqnarray}
\begin{eqnarray}
\Lambda_{2,3}^{(2)}=\beta_1\beta_2\frac{\left[\cosh\left(\frac{\Phi_0+\Phi_2}{2}\right)\cosh\left(\frac{\Phi_1-\Phi_3}{2}\right)+\cosh\left(\frac{\Phi_0-\Phi_2}{2}\right)\cosh\left(\frac{\Phi_1+\Phi_3}{2}\right)\right]}{\left[\cosh\left(\frac{\Phi_2-\Phi_3}{2}\right)\cosh\left(\frac{\Phi_1+\Phi_3}{2}\right)\beta_1^2-\cosh\left(\frac{\Phi_1-\Phi_3}{2}\right)\cosh\left(\frac{\Phi_2+\Phi_3}{2}\right)\beta_2^2\right]},\label{8}
\end{eqnarray}

Acting with $D_x$ in eqn. (\ref{Bx01})-(\ref{Bx23}) and using (\ref {fx01})- (\ref {fx23}), we find
\begin{eqnarray}
\partial_x(\Phi_0-\Phi_1)&=&\frac{4}{\beta_1^2}\sinh(\Phi_0+\Phi_1)+\frac{4i}{\beta_1}f_{0,1}D_x\left[\cosh\left(\frac{\Phi_0+\Phi_1}{2}\right)\right].\label{9a}\\
\partial_x(\Phi_1-\Phi_3)&=&\frac{4}{\beta_2^2}\sinh(\Phi_1+\Phi_3)+\frac{4i}{\beta_2}f_{1,3}D_x\left[\cosh\left(\frac{\Phi_1+\Phi_3}{2}\right)\right].\label{9b}\\
\partial_x(\Phi_0-\Phi_2)&=&\frac{4}{\beta_2^2}\sinh(\Phi_0+\Phi_2)+\frac{4i}{\beta_2}f_{0,2}D_x\left[\cosh\left(\frac{\Phi_0+\Phi_2}{2}\right)\right].\label{9c}\\
\partial_x(\Phi_2-\Phi_3)&=&\frac{4}{\beta_1^2}\sinh(\Phi_2+\Phi_3)+\frac{4i}{\beta_1}f_{2,3}D_x\left[\cosh\left(\frac{\Phi_2+\Phi_3}{2}\right)\right].\label{9d}
\end{eqnarray}
Notice that equations (\ref{9a})-(\ref{9d}) correspond formally to the pure bosonic case when the terms proportional to the fermionic superfields $f_{i,j}$ are neglected.
Equating the R.H.S. of the sum of eqns. (\ref{9a}) with (\ref{9b}) and (\ref{9c}) with (\ref{9d}) we find
\br
&&\frac{1}{\beta_1^2}\(\sinh(\Phi_0+\Phi_1) -\sinh(\Phi_2+\Phi_3)\)
+\frac{1}{\beta_2^2}\( \sinh(\Phi_1+\Phi_3) -\sinh(\Phi_0+\Phi_2)\)  \nonu \\
&=&-
\frac{i}{\beta_1}f_{0,1}D_x\left[\cosh\left(\frac{\Phi_0+\Phi_1}{2}\right)\right] -\frac{i}{\beta_2}f_{1,3}D_x\left[\cosh\left(\frac{\Phi_1+\Phi_3}{2}\right)\right]\nonu \\
&+&\frac{i}{\beta_2}f_{0,2}D_x\left[\cosh\left(\frac{\Phi_0+\Phi_2}{2}\right)\right]+\frac{i}{\beta_1}f_{2,3}D_x\left[\cosh\left(\frac{\Phi_2+\Phi_3}{2}\right)\right].
\label{11}
\er
Factorizing the L.H.S. of (\ref{11}) 
\br
&&\frac{1}{\beta_1^2}\(\sinh(\Phi_0+\Phi_1) -\sinh(\Phi_2+\Phi_3)\)
+\frac{1}{\beta_2^2}\( \sinh(\Phi_1+\Phi_3) -\sinh(\Phi_0+\Phi_2)\)  \nonu \\
&=& 2\cosh ({{\Phi_0+\Phi_1 +\Phi_2+\Phi_3}\o {2}})  \({{1}\o {\b_1^2}} \sinh ({{\Phi_0+\Phi_1 -\Phi_2-\Phi_3}\o {2}}) + {{1}\o {\b_2^2}} \sinh ({{-\Phi_0+\Phi_1 -\Phi_2+\Phi_3}\o {2}})\)\nonu 
\er
\br
= 2\cosh ({{\Phi_0+\Phi_1 +\Phi_2+\Phi_3}\o {2}})  \( \sinh ({{\Phi_0-\Phi_3}\o {2}}) \cosh ({{\Phi_1 -\Phi_2}\o {2}}) {{(\b_1^2 - \b_2^2)} \over {\b_1^2 \b_2^2}}\right. \nonu \\
 \left.+
\sinh ({{\Phi_1-\Phi_2}\o {2}}) \cosh ({{\Phi_0 -\Phi_3}\o {2}}) ({{{\b_1^2 + \b_2^2}\o {\b_1^2 \b_2^2}}}) \).
\label{12}
\er
 The vanishing of this expression leads to 
\br
\Phi_3 - \Phi_0 = 2 arctanh \(\d \; \tanh ({{\Phi_1-\Phi_2}\o {2}})\) \equiv \Gamma (\Phi_1-\Phi_2), \qquad \d = -\({{{\b_1^2 +\b_2^2}\o {\b_1^2 - \b_2^2}}}\).
\label{13}
\er
For the more general case taking into account the fermionic superfields $f_{i,j}$ we propose
   the following ansatz,
\begin{eqnarray}
\Phi_3=\Phi_0+\Gamma (\Phi_1-\Phi_2)+\Delta,
\label{Phi3}
\label{ansatz}
\end{eqnarray}
where $\Delta $ is a bosonic superfield proportional to the product $f_{0,1} f_{0,2}$, i.e.,
\br
 \D = \l f_{0,1} f_{0,2}, \qquad \l = \l(\Phi_1-\Phi_2).
\label{14}
\er
Due to the fact that $\Delta^2 = 0$, eqns.  (\ref{8})   take the  general form
\br
\Lambda_{1,3}^{(1)}&=&-a+c_1f_{0,1}f_{0,2},\nonumber\\
\Lambda_{1,3}^{(2)}&=&-b+c_2f_{0,1}f_{0,2},\nonumber\\
\Lambda_{2,3}^{(1)}&=&b+c_3f_{0,1}f_{0,2},\nonumber\\
\Lambda_{2,3}^{(2)}&=&a+c_4f_{0,1}f_{0,2},\label{15}
\end{eqnarray}
where $c_i = c_i (\Phi_0, \Phi_1, \Phi_2), i=1, \cdots 4$ do not contribute to eqns  (\ref{f13}) and (\ref{f23}). 
Substituting (\ref{ansatz}) and (\ref{14})  in (\ref{8}) we obtain
\begin{eqnarray}
a=\frac{\delta_1}{\sqrt{1-\delta^2\tanh^2\left(\frac{\Phi_1-\Phi_2}{2}\right)}},
\qquad
b=\frac{\delta\,
\mathrm{sech}\left(\frac{\Phi_1-\Phi_2}{2}\right)}{\sqrt{1-\delta^2\tanh^2\left(\frac{\Phi_1-\Phi_2}{2}\right)}}\label{16}
\end{eqnarray}
where $\delta_1=\frac{2\beta_1\beta_2}{(\beta_1^2-\beta_2^2)}$.
Inserting (\ref{15}) in (\ref{f13}) and (\ref{f23}) we find, since $f_{01}
^2 = f_{02}^2 =0$,
\begin{eqnarray}
f_{1,3}&=&-a f_{0,1}-b f_{0,2}, \nonumber\\
f_{2,3}&=&b f_{0,1}+a f_{0,2}.\label{fff}
\end{eqnarray}
From the fact that $\d^2 - \d_1^2 =1$, it follows that 
\begin{eqnarray}
f_{1,3}f_{2,3}=f_{0,1}f_{0,2}.\nonumber
\end{eqnarray}
Adding (\ref{Bx01}) and (\ref{Bx13}) we find
\begin{eqnarray}
D_x(\Phi_3-\Phi_0)=\frac{4i}{\beta_1}f_{0,1}\cosh\left(\frac{\Phi_0+\Phi_1}{2}\right)+\frac{4i}{\beta_2}f_{1,3}\cosh\left(\frac{\Phi_1+\Phi_3}{2}\right).\label{16}
\end{eqnarray}
Substituting (\ref{ansatz}) and (\ref{fff})  in (\ref{16}) we find 
\begin{eqnarray}
f_{0,1}\Sigma_1+f_{0,2}\Sigma_2+(D_x\lambda)
f_{0,1}f_{0,2}=0,\label{eq13}
\end{eqnarray}
where
\begin{eqnarray}
\Sigma_1&=&\partial_{\xi}\Gamma_{|_{\xi=(\Phi_1-\Phi_2)}}\frac{4i}{\beta_1}\cosh\left(\frac{\Phi_0+\Phi_1}{2}\right)-\lambda\frac{2i}{\beta_2}\sinh\left(\frac{\Phi_0+\Phi_2}{2}\right)\nonumber\\
&&-\frac{4i}{\beta_1}\cosh\left(\frac{\Phi_0+\Phi_1}{2}\right)-\Lambda_{1,3}^{(1)}\frac{4i}{\beta_1}\cosh\left(\frac{\Phi_0+\Phi_1+\Gamma}{2}\right),\nonumber\\
\Sigma_2&=&-\partial_{\xi}\Gamma_{|_{\xi=(\Phi_1-\Phi_2)}}\frac{4i}{\beta_2}\cosh\left(\frac{\Phi_0+\Phi_2}{2}\right)+\lambda\frac{2i}{\beta_1}\sinh\left(\frac{\Phi_0+\Phi_1}{2}\right)\nonumber\\
&&-\Lambda_{1,3}^{(2)}\frac{4i}{\beta_1}\cosh\left(\frac{\Phi_0+\Phi_1+\Gamma}{2}\right).\nonumber
\end{eqnarray}
The last term in (\ref{eq13}) vanishes since 
\begin{eqnarray}
D_x\lambda=\partial_{\xi}\lambda_{|_{\xi=(\Phi_1-\Phi_2)}}D_x(\Phi_1-\Phi_2), \nonumber
\end{eqnarray}
and $D_x(\Phi_1-\Phi_2)$, from (\ref{Bx01}) and (\ref{Bx02}),  is proportional to $f_{0,1}$ and $f_{0,2}$.  Since $f_{0,1}$ and $f_{0,2}$ are independent, (\ref{eq13}) yields a pair of algebraic equations for $\l$, i.e. $\Sigma_1= \Sigma_2 = 0$  which are satisfied by
\begin{eqnarray}
\lambda=-\frac{4\sinh\left(\frac{\Phi_1-\Phi_2}{2}\right)\beta_1\beta_2(\beta_1^2+\beta_2^2)}{\beta_1^4+\beta_2^4-2\cosh(\Phi_1-\Phi_2)\beta_1^2\beta_2^2}.\label{lambda}
\end{eqnarray}
and therefore  
\begin{eqnarray}
\Phi_3=\Phi_0+2\,\mathrm{Arctanh}\left[\left(\frac{\beta_2^2+\beta_1^2}{\beta_2^2-\beta_1^2}\right)\tanh\left(\frac{\Phi_1-\Phi_2}{2}\right)e^{\Omega
f_{0,1}f_{0,2}}\right], \label{solution}
\end{eqnarray}
where
\begin{eqnarray}
\Omega=\d_1 \mathrm{sech}\left(\frac{\Phi_1-\Phi_2}{2}\right).\nonumber
\end{eqnarray}
In order to write eqn. (\ref{solution}) in components, we need to specify the superfields $f_{0,1}, f_{0,2}$   
in terms of the components of $\Phi_0, \Phi_1, \Phi_2$.  These are given by eqns. (5.91), (5.94), (5.96) and (5.98) in  the  the appendix of  ref. \cite{ymai}. 
Introducing $\sigma_k=-\frac{2}{\beta_k^2}$ $(k=1,2)$, the solution (\ref{solution}) in components according to (\ref{2}) becomes
\begin{eqnarray}
\phi_3&=&\phi_0 +
2\,\mathrm{Arctanh}\left[\delta\tanh\left(\frac{\phi_1-\phi_2}{2}\right)\right]\nonumber\\
&&-\frac{\Delta_2}{8\sqrt{\sigma_1\sigma_2}}\left[\frac{\bar{\psi}_0(\bar{\psi}_1-\bar{\psi}_2)+\bar{\psi}_1\bar{\psi}_2}{\cosh\left(\frac{\phi_0+\phi_1}{2}\right)\cosh\left(\frac{\phi_0+\phi_2}{2}\right)}\right],\label{14a}\\
\bar{\psi}_3&=&\bar{\psi}_0+\Delta_1(\bar{\psi}_1-\bar{\psi}_2)\nonu\\
&&-\frac{\Delta_2}{2}\left[\sqrt{\frac{\sigma_2}{\sigma_1}}\frac{\sinh\left(\frac{\phi_0+\phi_2}{2}\right)}{\cosh\left(\frac{\phi_0+\phi_1}{2}\right)}(\bar{\psi}_0-\bar{\psi}_1)-\sqrt{\frac{\sigma_1}{\sigma_2}}\frac{\sinh\left(\frac{\phi_0+\phi_1}{2}\right)}{\cosh\left(\frac{\phi_0+\phi_2}{2}\right)}(\bar{\psi}_0-\bar{\psi}_2)\right],\label{14b}\\
\psi_3&=&\psi_0+\Delta_1(\psi_1-\psi_2)\nonu \\
&&-\frac{\Delta_2}{2}\left[\sqrt{\frac{\sigma_2}{\sigma_1}}\frac{\sinh\left(\frac{\phi_0-\phi_1}{2}\right)}{\cosh\left(\frac{\phi_0-\phi_2}{2}\right)}(\psi_0+\psi_2)-\sqrt{\frac{\sigma_1}{\sigma_2}}\frac{\sinh\left(\frac{\phi_0-\phi_2}{2}\right)}{\cosh\left(\frac{\phi_0-\phi_1}{2}\right)}(\psi_0+\psi_1)\right],\label{14c}
\end{eqnarray}
where
\begin{eqnarray}
\Delta_1&=&\frac{2}{\sinh\left(\phi_1-\phi_2\right)}\left[\frac{\delta\tanh\left(\frac{\phi_1-\phi_2}{2}\right)}{1-\delta^2\tanh^2\left(\frac{\phi_1-\phi_2}{2}\right)}\right],\nonumber\\
\Delta_2&=&\frac{A\sinh\left(\frac{\phi_1-\phi_2}{2}\right)}{B-\sinh^2\left(\frac{\phi_1-\phi_2}{2}\right)},\nonumber\\
\delta&=&\frac{\sigma_1+\sigma_2}{\sigma_1-\sigma_2}, \qquad
A=\frac{\sigma_1+\sigma_2}{\sqrt{\sigma_1\sigma_2}}, \qquad
B=\frac{(\sigma_1-\sigma_2)^2}{4\,\sigma_1\sigma_2}.\label{15a}
\end{eqnarray}

{\it One soliton Solution}

The  Backlund equations in components for  $\Phi_0=0$ take the form )from (\ref{6}):
\begin{eqnarray}
\partial_x\phi_1&=&2\,\sigma_1\sinh\phi_1,\label{ex1} \qquad 
\partial_t\phi_1=\frac{2}{\sigma_1}\sinh\phi_1,\label{ex2}\\
\bar{\psi}_1&=&2\sqrt{2 \sigma_1} \cosh \( \phi_1 /2\)f_1^{(0,1)}, \qquad  \psi_1 = 2 \sqrt{ {{2}\over {\s_1}}} \cosh \(\phi_1 /2\)f_1^{(0,1)}
\label{ex3} \\
\pa_x f_1^{(0,1)} &=& \sqrt{{{\s_1}\over {2}}} \cosh \(\phi_1 /2 \) \bar \psi_1, \qquad \pa_t f_1^{(0,1)} = {{1}\over {\sqrt{2\s_1}}} \cosh \(\phi_1 /2\) \psi_1
\end{eqnarray}
By direct integration we obtain
\begin{eqnarray}
\phi_1&=&\ln\left(\frac{1+E_1}{1-E_1}\right), \qquad
E_1=b_1\exp\left(2\sigma_1x+2\sigma_1^{-1}t\right),\label{ex4}\\
\bar{\psi}_1&=&\epsilon_1\frac{a_1}{b_1}E_1\left(\frac{1}{1+E_1}+\frac{1}{1-E_1}\right),
\qquad \psi_1=\frac{\bar{\psi}_1}{\sigma_1},\label{ex5}
\end{eqnarray}
where $a_1$ and
 $b_1$ are arbitrary constants and  $\epsilon_1$  is a fermionic parameter.

{\it Two soliton Solution}

Choosing   $\Phi_0=0$ and $\Phi_1, \Phi_2$ as one soliton solutions with components
\begin{eqnarray}
\phi_k&=&\ln\left(\frac{1+E_k}{1-E_k}\right), \qquad
E_k=b_k\exp\left(2\sigma_kx+2\sigma_k^{-1}t\right),\nonumber\\
\bar{\psi}_k&=&\epsilon_k\frac{a_k}{b_k}E_k\left(\frac{1}{1+E_k}+\frac{1}{1-E_k}\right),\nonumber\\
\psi_k&=&\frac{\bar{\psi}_k}{\sigma_k}, \qquad k=1,2 \nonumber
\end{eqnarray}
where $a_k$, $b_k$ are arbitrary constants and $\epsilon_k$
fermi\^onic parameters we find from 
 (\ref{solution}), 
\begin{eqnarray}
\phi_3&=&2\,\mathrm{Arctanh}\left[\delta\tanh\left(\frac{\phi_1-\phi_2}{2}\right)\right]-\frac{\Delta_2}{8\sqrt{\sigma_1\sigma_2}}\left[\frac{\bar{\psi}_1\bar{\psi}_2}{\cosh\left(\frac{\phi_1}{2}\right)\cosh\left(\frac{\phi_2}{2}\right)}\right],\label{ex4}\\
\bar{\psi}_3&=&\left[\Delta_1+\frac{\Delta_2}{2}\sqrt{\frac{\sigma_2}{\sigma_1}}\frac{\sinh\left(\frac{\phi_2}{2}\right)}{\cosh\left(\frac{\phi_1}{2}\right)}\right]\bar{\psi}_1-\left[\Delta_1+\frac{\Delta_2}{2}\sqrt{\frac{\sigma_1}{\sigma_2}}\frac{\sinh\left(\frac{\phi_1}{2}\right)}{\cosh\left(\frac{\phi_2}{2}\right)}\right]\bar{\psi}_2,\label{ex5}\\
\psi_3&=&\left[\Delta_1-\frac{\Delta_2}{2}\sqrt{\frac{\sigma_1}{\sigma_2}}\frac{\sinh\left(\frac{\phi_2}{2}\right)}{\cosh\left(\frac{\phi_1}{2}\right)}\right]\psi_1-\left[\Delta_1-\frac{\Delta_2}{2}\sqrt{\frac{\sigma_2}{\sigma_1}}\frac{\sinh\left(\frac{\phi_1}{2}\right)}{\cosh\left(\frac{\phi_2}{2}\right)}\right]\psi_2,\label{ex6}
\end{eqnarray}

By rescaling  of parameters 
\begin{eqnarray}
&&\sigma_k\to \gamma_k , \qquad \epsilon_k\to c_k \qquad k=1,2\nonumber\\
&&b_1\to
\frac{b_1}{2}\left(\frac{\gamma_1-\gamma_2}{\gamma_1+\gamma_2}\right),
\qquad b_2\to
-\frac{b_2}{2}\left(\frac{\gamma_1-\gamma_2}{\gamma_1+\gamma_2}\right),\nonumber\\
&&a_1\to
-\gamma_1\left(\frac{\gamma_1-\gamma_2}{\gamma_1+\gamma_2}\right),
\qquad a_2\to
\gamma_2\left(\frac{\gamma_1-\gamma_2}{\gamma_1+\gamma_2}\right).\nonumber
\end{eqnarray}
 we verify that solution (\ref{ex4}) and  (\ref{ex5}) coincide precisely with solution 
(3.28)-(3.29) of ref. \cite{ymaipla} (after choosing  $\gamma_3=-\gamma_1$ and $\gamma_4=-\gamma_2$).

\vskip .5cm

 \noindent
{\bf Acknowledgements} \\
\vskip .1cm \noindent
{  LHY acknowledges support from Fapesp, JFG and AHZ thank CNPq for partial support.}

\end{document}